\documentclass[%
 reprint,
 amsmath,amssymb,
prb,
]{revtex4-1}

\usepackage{graphicx}% Include figure files
\usepackage{dcolumn}% Align table columns on decimal point
\usepackage{bm}% bold math
\usepackage{color}
\usepackage[english]{babel}
\begin{document}

\preprint{APS/123-QED}

\title{First-principles calculations of thermal electron emission from H$^{-}$ in silicon}

\author{Yoshiyuki Yamamoto}
    \altaffiliation[Current affiliation:]{Research and Services Division of Materials Data and Integrated System, 
    National Institute for Materials Science, 1-1 Namiki, Tsukuba, Ibaraki 305-0044, Japan}
    \email[E-mail:]{YAMAMOTO.Yoshiyuki@nims.go.jp}
    \affiliation{The Institute for Solid State Physics, The University of Tokyo, 5-1-5 Kashiwanoha, Kashiwa, Chiba 277-8581, Japan}
\author{Alpin Novianus Tatan}
    \affiliation{The Institute for Solid State Physics, The University of Tokyo, 5-1-5 Kashiwanoha, Kashiwa, Chiba 277-8581, Japan}
\author{Osamu Sugino}
    \affiliation{The Institute for Solid State Physics, The University of Tokyo, 5-1-5 Kashiwanoha, Kashiwa, Chiba 277-8581, Japan}

\date{\today}% It is always \today, today,

\begin{abstract}
Thermal electron emission process of a hydrogen impurity is an important topic of fundamental semiconductor physics. 
Despite of decades-long  study, theory is not established yet.
Here, we study the process of $\mathrm{H}^{-}$ in silicon, $\mathrm{H^{-}} \to \mathrm{H^{0}} + e^{-}$, 
using a first-principles calculation.
Our calculation indicates that the process consists of two steps:
slow diffusion of H$^{-}$ from a tetrahedral site to a bond-center site, which is the rate-limiting step,
and faster nonradiative transition from H$^{-}$ to $\mathrm{H}^{0} + e^{-}$ that occurs subsequently at the body-center site. 
The calculated rate is consistent with a deep level transient spectroscopy experiment.

\end{abstract}

\maketitle

\section{Introduction}
Due to its amphoteric nature and fast diffusivity, hydrogen impurities passivate various defects 
and help increase the efficiency of various semiconductor devices. 
Their nonradiative carrier capture/emission capability \cite{alkauskas_tutorial:_2016} 
significantly affects diffusivity and carrier lifetimes.
Due to its technological importance, this phenomenon has been studied over the past decades.

In silicon, hydrogen is known to exist as H$^{+}$ or H$^{-}$ depending on 
doping condition\cite{tavendale_field_1985, johnson_interstitial_1986}, 
while H$^{0}$ is metastable\cite{johnson_inverted_1994} 
and can exist only at very low temperatures\cite{bonde_nielsen_bond-centered_1999}
or under illumination\cite{gorelkinskii_electron_nodate}.
These properties can be consistently explained by first-principles total energy 
calculations\cite{van_de_walle_theory_1989, herring_energy_2001}, 
which indicate that H$^{+}$ or H$^{-}$ is thermodynamically stable
while H$^{0}$ is always unstable against the reaction $2\mathrm{H}^{0} \to \mathrm{H}^{+} + \mathrm{H}^{-}$.
The transition between different charge states was experimentally studied 
with a deep level transient spectroscopy (DLTS)\cite{herring_energy_2001}.
The DLTS experiment showed that 
%the reaction $\mathrm{H}^{0} \to \mathrm{H}^{+} + e^{-}$ is fast, 
thermal electron emission from H$^{-}$, {\it i.e.}, $\mathrm{H}^{-} \to \mathrm{H}^{0} + e^{-}$,
has a slow reaction rate, $r = 2.8 \times 10^{-1} \, \mathrm{s}^{-1}$ 
at room temperature with activation energy $E_{\mathrm{a}} = 0.84 \, \mathrm{eV}$.
%The difference in the rate has been understood 
The activation energy is usually thought 
to correspond to the energy difference between H$^{-}$ and H$^{0} + e^{-}$.
However, density functional theory (DFT) calculation of this quantity only amounts to 0.50 eV. 
The nature of the observed activation energy in the experiment therefore remains to be investigated.

Here, we show our theoretical study on the rate for thermal electron emission by H$^{-}$ in silicon.
Although there are many studies treating hydrogen impurity in silicon,
as far as we know, there has been no first-principles study on the rate 
because it requires accurate description of the band gap, the defect thermodynamics, and electron-phonon couplings.
Recently, however, Kim {\it et al.} demonstrated that first-principles calculation can reproduce
the rate for nonradiative carrier capture from the $DX$ center in GaAs 
despite large anhnarmonicity of the potential energy surface (PES)\cite{kim_anharmonic_2019}.
Because the PES for hydrogen impurity in silicon is also known to be anharmonic\cite{herring_energy_2001},
their approach should help us to elucidate the reason for the high activation energy observed in the DLTS experiment on H$^{-}$.

In this context, we calculated thermodynamic transition level with screened-hybrid DFT and confirmed that
the calculated result is different from the experimental activation energy.
Then, we constructed a configuration coordinate diagram 
for the transition between $\mathrm{H}^{-}$ and $\mathrm{H}^{0} + e^{-}$.
The calculated diagram for H$^{-}$ was found to have local minimum not only
at its most stable tetrahedral site but also at a bond center site, 
which is the global minimum for $\mathrm{H}^{0} + e^{-}$.
The existence of local minimum at the bond-center site is a key
to explain the fast thermal electron emission because nuclear wavefunctions overlap considerably at the bond-center site.
The calculated rate shows that, with the assumption of thermal equilibrium,
thermal electron emission is much faster than the experiment,
indicating another step should rate-determine the whole process.
Finally, we calculated the migration barrier of H$^{-}$ between tetrahedral and bond-center site,
and conclude that the migration is the rate-limiting step for thermal electron emission from H$^{-}$.

\section{Methods}
We performed first-principles calculations based on DFT
\cite{hohenberg_inhomogeneous_1964, kohn_self-consistent_1965}
with Vienna Ab initio Simulation Package (\textsc{vasp})\cite{kresse_efficient_1996, kresse_efficiency_1996, kresse_ab_1995, kresse_ab_1994}, 
where the Kohn-Sham orbitals are expressed by the plane-wave basis.
We used the screened hybrid functional formulated as 
Heyd-Scuseria-Ernzerhof exchange correlation functional (HSE06 functional)\cite{krukau_influence_2006}.
Projector augmented wave (PAW) method was used for the interactions between valence electrons and ions.
The cutoff energy for the plane-wave basis was $400 \, \mathrm{eV}$, 
which was found to give sufficient accuracy for defect thermodynamics and electron-phonon coupling in our system.
We optimized the lattice constant of silicon with conventional cell calculations 
using $\Gamma$-centered $6 \times 6 \times 6$ k-point mesh for Brillouin-zone integration.
The calculated lattice constant is 5.433 \AA, in good agreement with experimental one 
extrapolated to $0\, \mathrm{K}$, 5.430 \AA\cite{staroverov_tests_2004}.

For defect calculations, we use $3 \times 3\times 3$ 
super cell of conventional cell, including 216 Si atoms,
in order to suppress spurious interactions between defects.
To reduce the computational cost, $\Gamma$-point calculation was performed.
The calculated band gap of silicon is 1.31 eV which is larger than the experimental one, 1.17 eV, 
because the conduction band bottom is not correctly sampled with $\Gamma$-point in $3\times 3\times 3$ super cell.
To obtain correct energetics between H$^{-}$ and H$^{0}$ + $e^{-}$,
we adjust the Kohn-Sham energies of the conduction bands by shifting them 
by $-0.14 \, \mathrm{eV}$ in the calculated configuration coordinate diagram.
Finite-size correction for charged defect is calculated
with the Markov-Payne correction\cite{makov_periodic_1995}
by extrapolating the calculated energies 
with $2\times2\times3$, $2\times3\times3$ and $3\times3\times3$ supercells to infinite cell size.
For the extrapolation, we used $2 \times 2 \times 2$ k-point for all supercells:
this k-point mesh is found to be necessary for the extrapolation to work well.
The correction was calculated for H$^{-}$ at a tetrahedral site,
and the same correction was used for all other configurations.
The finite-size correction changes
the calculated thermodynamic transition level, $\varepsilon (0/-)$, by +0.06 eV.

The rate for thermal electron emission was calculated
with the formalism proposed by Alkauskas {\it et al}\cite{alkauskas_first-principles_2014}.
In the formalism, with the aid of the static approximation for electronic states 
and one-dimensional approximation for nuclear degrees of freedom,
the rate for nonadiabatic transition from electronic state $i$ to $j$ with multiphonon absorption 
is calculated with the Fermi's golden rule:
\begin{equation}
    r_{ij} = \frac{2\pi}{\hbar} |W_{ij}|^2 \sum_{n} \omega_n \sum_{m} 
    |\langle \chi_{jm} |Q-Q_0| \chi_{in} \rangle|^2 \delta(E_{in} - E_{jm}).
    \label{eq: rate}
\end{equation}
$W_{ij}$ is the electron matrix element defined as $W_{ij} =  \langle \psi_j | \frac{\partial H}{\partial Q}| \psi_i \rangle$,
where $|\psi_{i(j)} \rangle$ is the Kohn-Sham orbital for state $i(j)$.
With first-order perturbation theory, $W_{ij}$ can be expressed as 
$W_{ij} = (\varepsilon_j - \varepsilon_i) \langle \psi _j | \frac{\partial \psi_i }{\partial Q}\rangle$,
where $\varepsilon_{i}$ is the Kohn-Sham energy for a state $i$.
The Kohn-Sham orbitals overlaps, $\langle \psi _j | \frac{\partial \psi_i }{\partial Q}\rangle$, is calculated
with all-electron wavefunctions including PAW augmented core contributions reproduced by \textsc{pawpyseed} code\cite{bystrom_pawpyseed:_2019}.
$\omega_n$ is the thermal weight for the n-th eigenstate of nuclear wavefunction, $|\chi_{in}\rangle$.
The nuclear wave functions and eigenenergies are calculated by solving 
the one-dimensional Schr\"{o}dinger equation for configuration coordinate diagram
to account for large anharmonicity of the potential energy surface\cite{kim_anharmonic_2019}.
$Q$ is the configuration coordinate defined as $Q = \sqrt{M} R$, 
where M is a diagonal matrix with masses in its diagonal element
and R is the $3N$-dimensional coordinate for lattice.
We assume a linear reaction pathway between the most stable configurations of two charge states, H$^{0}$ and H$^{-}$,
following the scheme proposed by Alkauskas {\it et al.}\cite{alkauskas_first-principles_2014}
although this treatment is known to underestimate the rate\cite{shi_comparative_2015}
because the linear pathway does not necessarily give a major contribution.
The electron emission rate from H$^{-}$ is calculated by summing up the contributions 
of the transition from a defect state to conduction band states.
The electron emission rate is thus calculated as $r_{i} = \sum_{j=\mathrm{CB}} r_{ij}$,
where $i$ is the defect state, and j is the conduction band state.
As the summation for the conduction bands, 
we take the bands within $0.15 \, \mathrm{eV}$ above the conduction band minimum.
We have confirmed that conduction bands above them give only negligible effects to the calculated rate.

The delta function in Eq.~(\ref{eq: rate}) is approximated as Gaussian with finite width, $\sigma$,
which corresponds to a lifetime of the vibrational mode considered in the configuration coordinate.
Although lifetimes of local vibrational modes at low temperatures are known\cite{budde_vibrational_2000},
we cannot use those values as $\sigma$ because the local vibrational modes are different from our reaction pathway.
Thus, we tried several values of $\sigma$ to investigate its effect on the calculated reaction rate.
The value of $\sigma$ is, indeed, found to affect the calculated rate.
However, our conclusion is robust against the choice of $\sigma$ value,
and we used $\sigma = 0.026 \, \mathrm{eV}$ throughout this study.

\section{Results}
\subsection{Thermodynamics}
Firstly, we investigate the thermodynamics of hydrogen in silicon.
The calculated energies suggest that the most stable site is the bond-center (BC) site for H$^{0}$ while
the tetrahedral (Td) site for H$^{-}$, 
and the calculated thermodynamic transition level is $\varepsilon(0/-) = 0.67 \, \mathrm{eV}$ from the valence band maximum.
These results are in good agreement with previous theoretical study using local density functional\cite{herring_energy_2001}.
For the electron emission process, the energy difference between $\mathrm{H}^{0} + e^{-}$ and $\mathrm{H}^{-}$
is calculated to be $E_{g} - \varepsilon(0/-) = 0.50 \, \mathrm{eV}$, where $E_{g}$ is the band gap of silicon.
Thus, the activation energy for thermal electron emission is expected to be $0.50 \, \mathrm{eV}$ from first-principles calculations.
This value is, however, inconsistent with an experimental activation energy, $0.84 \, \mathrm{eV}$, 
obtained by a DLTS experiment\cite{herring_energy_2001}.
Considering the inconsistency, we investigated possible origins of the discrepancy from computational point of views.
We calculated the effects of thermal expansion of silicon at room temperature, zero-point energy of hydrogen and 
the type of pseudopotentials (hard/standard).
These effects change the calculated $\varepsilon(0/-)$ as 
$-0.01 \, \mathrm{eV}$, $+0.06 \, \mathrm{eV}$, and $-0.02 \, \mathrm{eV}$, respectively.
None of these effects is thus confirmed to fill the gap between theoretical and experimental values.
In conclusion, our calculation indicates that 
thermodynamics could not explain the activation energy observed in the DLTS experiment.

\subsection{Configuration Coordinate Diagrams}
\begin{figure}
\includegraphics[width=8.6cm]{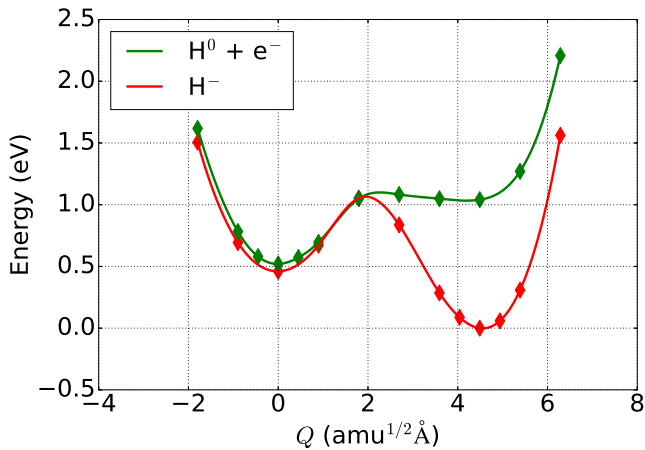}                                                                            
\caption{\label{fig:pes} 
(Color online) Configuration coordinate diagrams of $\mathrm{H}^{0} + e^{-}$ and H$^{-}$.
For $\mathrm{H}^{0} + e^{-}$, electron is assumed to occupy the conduction band bottom.
The dot symbols show the calculated energies from first principles, 
and the lines show the ones calculated with the spline interpolation between the calculated ones.
$Q=0$ amu$^{1/2}$ \AA \, is the most stable configuration for H$^{0}$ (BC site)
while the configuration for $Q=4.49$ amu$^{1/2}$ \AA \, is the most stable one for H$^{-}$ (Td site).}
\end{figure}

Figure \ref{fig:pes} shows the configuration coordinate diagram for the transition between 
$\mathrm{H}^{-}$ and $\mathrm{H}^{0}+e^{-}$. 
The configuration coordinate is taken as the linear interpolation between
their most stable Td site for H$^{-}$ and BC site for H$^{0}$.
The diagram for $\mathrm{H}^{0} + e^{-}$ is calculated as the summation of
total energy of $\mathrm{H}^{0}$ and the Kohn-Sham energy of the conduction band minimum of pristine silicon.
%This treatment could be justified in dilute limit of defect
%because the Kohn-Sham energy should not be affected by defect in the dilute limit.
The calculated configuration coordinate diagrams have two unique features:
firstly, the H$^{0}$ state shows large anharmonicity around the configurations of the Td site.
At the configuration of the Td site, the hydrogen occupies the interstitial void of silicon.
Because the radius of H$^{0}$ is smaller than that of H$^{-}$,
the H$^{0}$ interacts less strongly with Si.
Thus, the flat PES appears around the Td site for H$^{0}$
while the PES of H$^{-}$ has a deep minimum at the Td site.
%Due the flatness of the PES, the energy difference of
The calculated energy difference of $\mathrm{H}^{0} + e^{-}$ and $\mathrm{H}^{-}$ 
around Td site is large, $1.0 \, \mathrm{eV}$.
The large energy gap indicates that 
the activation energy of 1.0 eV is required for H$^{-}$ to emit an electron around Td site.
Secondly, H$^{-}$ state has a local minimum at the BC site configuration.
The existence of local minimum has been confirmed by additional structural optimization calculation.
Because both $\mathrm{H}^{-}$ and $\mathrm{H}^{0} + e^{-}$ have local minimum at the same configuration,
the nuclear wavefunctions should overlap significantly.
The electron emission rate is, therefore, expected to be large at the BC site.
We should note that the local minimum at the BC site for H$^{-}$
was not observed in the work by Herring {\it et al}\cite{herring_energy_2001}. 
The reason for the discrepancy is likely to be the difference in the configuration coordinates or
exchange-correlation functional used to calculate the PES,
{\it i.e.,} HSE06 functional in our calculation and local density functional in their work.

\begin{figure}
\includegraphics[width=8.6cm]{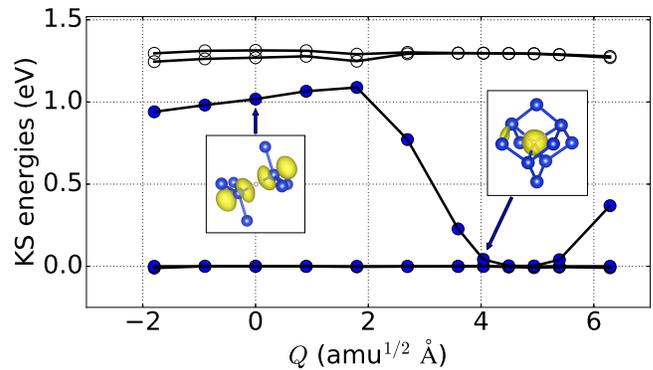}
\caption{\label{fig:eigens}
(Color online) 
The calculated Kohn-Sham energies of H$^{-}$ for the bands around defect level along the configuration coordinate.
Left and right inset figures show the partial charge densities of defect state 
with H at the BC site and near the Td site, respectively.
In the insets, large blue spheres correspond to Si atoms and small pink spheres to H atom.
Filled and open circles correspond to occupied and unoccupied states, respectively.
Configuration coordinate is same as Figure \ref{fig:pes}.}
\end{figure}

Figure \ref{fig:eigens} shows the Kohn-Sham energies of valence bands, 
a defect state, and conduction bands along the configuration coordinate.
The hydrogen insertion into the BC site makes the anti-bonding state of Si-Si bonding stable
due to expansion of Si-Si bond and attractive interaction by proton.
Thus, the anti-bonding state appears in the band gap as a defect state.
As the configuration changes from that of BC site to Td site,
the anti-bonding state disappears while another in-gap state appears.
The new in-gap state corresponds to the defect state at Td site,
which is the electronic state localized around the hydrogen in the interstitial void of Si.
The defect levels of H$^{-}$, however, 
do not exist in the band gap at the Td site configuration % ($Q=4.49$ amu$^{1/2}$ \AA)
as shown in the Figure \ref{fig:eigens}.
This apparently causes a problem in applying the Fermi's golden rule formalism
because the formalism requires to identify the initial defect state but the state 
is hybridized with the valence band.
It does not, however, hamper the calculation of nonradiative electron emission from H$^{-}$.
Because, as discussed in the previous paragraph, 
the emission is likely to be dominated at the BC site
where the energy required to activate the emission is much smaller than
that at the Td site.
%We circumvent this difficulty by considering the fact that 
%low energy transition is likely to be dominant around BC site 
%because nuclear wave functions of H$^{0}$+e$^{-}$ state and $H^{-}$ state could overlap
%at the energy 0.5 eV above the energy minimum 
%while it could occur at least 1.0 eV above the energy minimum around Td site.
Therefore, we can use the Kohn-Sham orbitals around BC site to calculate
$W_{ij}$ in Eq\@. (\ref{eq: rate}) to discuss the electron emission at room temperature.
It should be noted that we neglect relative efficiency of 
the Td-to-BC migration of H$^{-}$ and the thermal electron emission at the BC site.
Hence the migration is assumed to occur quickly toward thermal equilibrium so that
the thermal weight, $\omega_n$, is allowed to be used in Eq\@. \ref{eq: rate}.
%That is, the population of the initial state $|\chi_{in} \rangle$ is determined
%regardless of whether they exist around Td site or BC site.
%The population of hydrogen in the local minima around BC site
%is assumed to be determined by the thermodynamics.
Later, we will discuss the validity of this assumption
based on the calculated rate from first principles.

\subsection{Nonradiative Electron Emission}
The electronic matrix element, $W_{ij}$, is calculated with the finite-displacement method
at the configuration of the BC site as discussed in the previous section.
In table \ref{tab:Wij}, we tabulated the values of $W_{ij}$ 
which give major contributions to the calculated rates.
\begin{table}[b]
\caption{\label{tab:Wij}
The calculated values of $W_{ij}$ for H$^{-}$
between a defect state, $i$, and conduction band states, $\{ j \}$, at the BC site configuration. 
$\varepsilon_j$ is the Kohn-Sham energy for state $j$.
Only the $W_{ij}$'s that have major contributions to the electron emission rate
are tabulated.}
\begin{ruledtabular}
\begin{tabular}{lcc}
\textrm{$j$}&
\textrm{$\varepsilon_j - \varepsilon_{\mathrm{CBM}}$} (eV)&
$W_{ij} (\mathrm{eV}/\mathrm{amu^{1/2}}$  \AA)\\
\colrule
CBM    & 0.000  & $2.34\times10^{-2}$ \\
CBM+2  & 0.042 & $6.47\times10^{-2}$ \\
CBM+5  & 0.047 & $7.66\times10^{-2}$ \\
CBM+9  & 0.102 & $9.80\times10^{-2}$ \\
CBM+10 & 0.102 & $2.39\times10^{-2}$ \\
\end{tabular}
\end{ruledtabular}
\end{table}
As shown in the table, not only the CBM
but also other bands within $0.15 \ \mathrm{eV}$ from CBM have large values of $W_{ij}$,
indicating the importance to include those contributions into the calculations.
Only eleven conduction bands are included in the calculations 
because the other conduction bands are well-seperated by $0.5 \ \mathrm{eV}$ and are thus negligible.
The nuclear wavefunctions in Eq\@. (\ref{eq: rate}) are obtained by solving the one-body Schr\"{o}dinger equation 
with PES obtained with spline interpolation of the calculated PES.
%In figure \ref{fig:pes}, the interpolated PES for H$^{-}$ and H$^{0} + e^{-}\mathrm{(CBM)}$ are shown.
%For electron emission to a conduction band other than CBM,
%the PES of $\mathrm{H}^{0} + e^{-}$ is shited
%To calculate the rate for a transition from the defect state to a conduction band, $j$, 
%the PES for $\mathrm{H}^{0} + e^{-}$ is shifted according to the Kohn-Sham energy of the conduction band
%in order to account for the energy difference between $\varepsilon_j$ and $\varepsilon_{\mathrm{CBM}}$.
%The difference of Kohn-Sham energies are tabulated in Table \ref{tab:Wij}.
We note that the configuration coordinate diagrams around the BC site
is in the Marcus inverted region\cite{marcus_electron_1993}:
because positions of the local minima are virtually the same within our configuration coordinate,
the overlaps of nuclear wavefunctions are expected to be large compared to that in the normal region,
thus possibly giving large rate for thermal electron emission.

\begin{figure}
\includegraphics[width=8.6cm]{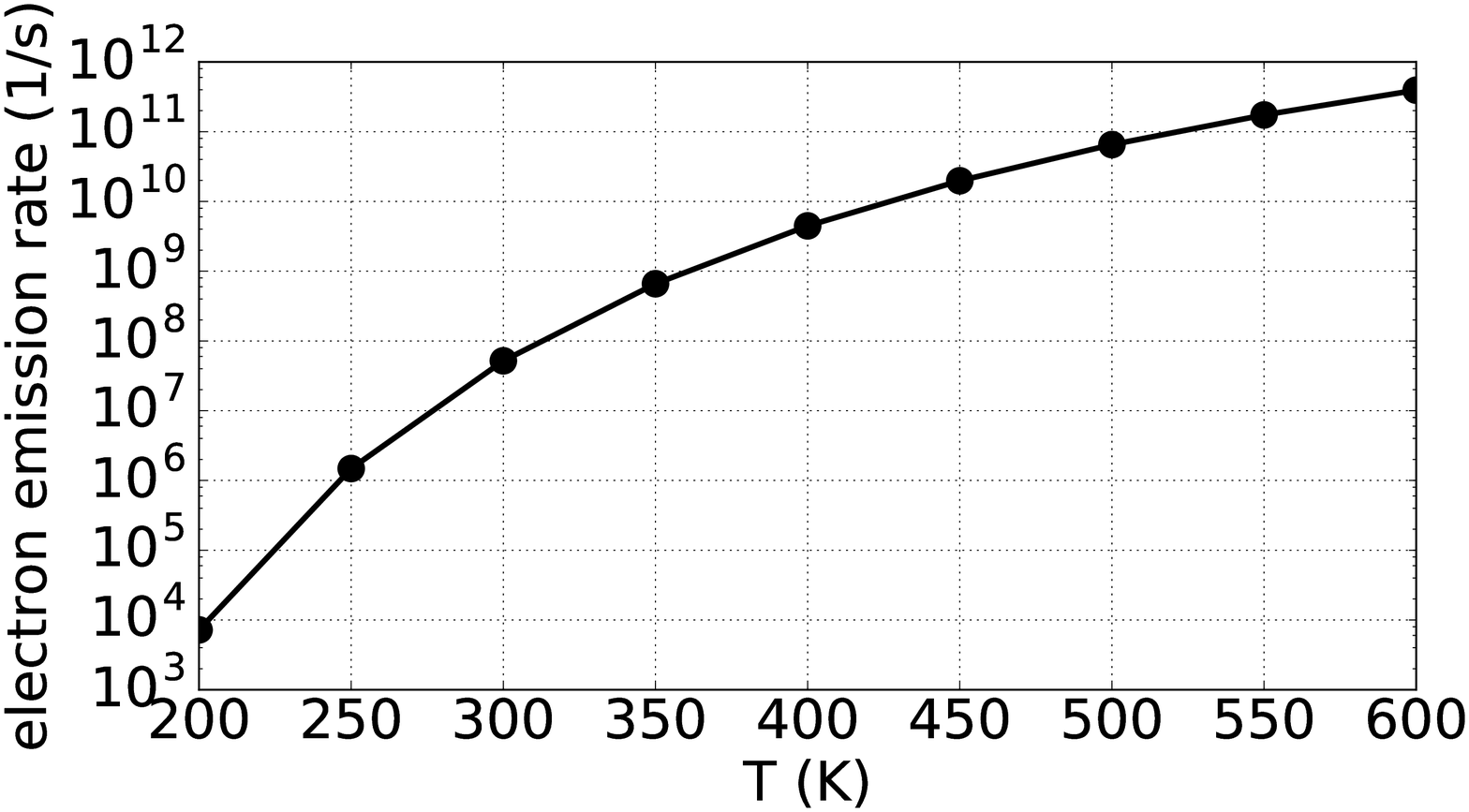}
\caption{\label{fig:rates}
The calculated rate for electron emission from H$^{-}$.
The black circles are the calculated values and the black line is guide for eyes.
Activation energy, $E_{a}$, is estimated to be $0.46 \, \mathrm{eV}$ 
by fitting the calculated rates to the Arrhenius plot, $r = r_{0}\exp(-E_{a}/kT)$}.
\end{figure}

Figure \ref{fig:rates} shows the calculated rate for the thermal electron emission from H$^{-}$ to conduction bands.
The calculated rate at room temperature is $5.2 \times 10^{7} \, \mathrm{s}^{-1}$
and the activation energy is estimated to be $0.46 \, \mathrm{eV}$ from the Arrhenius plot.
The calculated activation energy is consistent with the results expected with thermodynamics calculations, 
$E_g - \varepsilon(0/-) = 0.50 \, \mathrm{eV}$.
Although the calculated rates depend on the smearing factor $\sigma$ 
to describe the $\delta$ function, the calculated rates are much larger than the experimental value, 
$r = 2.8 \times 10^{-1} \, \mathrm{s}^{-1}$ at room temperature\cite{herring_energy_2001}.
%This fact suggests that the rate-determining step is not 
%the electron emission step at the BC site: 
%there is likely to exist another step which is responsible for the electron emission rate.

\begin{figure}
\includegraphics[width=8.6cm]{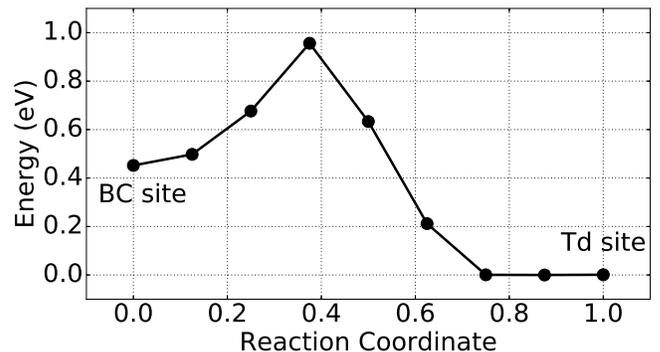}
\caption{\label{fig:cineb}
The minimum energy path for H$^{-}$ migration calculated with the climbing image nudged elastic band method.
The calculated activation energy is 0.96 eV from the Td site.}
\end{figure}

The argument so far assumes thermal equilibrium:
{\it i.e.} the diffusion of $\mathrm{H}^{-}$ from the Td site to the BC site 
is assumed to be much faster than the thermal electron emission at BC site,
thus enabling the use of thermal weight, $\omega_n$ in Eq\@. (\ref{eq: rate}) as a population of initial state.
To test this assumption, 
we calculate the activation energy for the migration of H$^{-}$
with the climbing image nudged elastic band method\cite{henkelman_improved_2000, henkelman_climbing_2000}.
Figure \ref{fig:cineb} shows the calculated PES for H$^{-}$ migration between Td site and BC site.
Here, the configuration of the BC site is optimized for H$^{-}$,
thus slightly different from the one used in the configuration coordinate diagram in Figure \ref{fig:pes},
which is the configuration optimized for H$^{0}$.
The calculated activation energy is $0.96 \, \mathrm{eV}$, 
which is comparable to the one from the DLTS experiment, $0.84 \, \mathrm{eV}$.
The calculated activation energy is used to estimate 
the rate for the migration with a simple formula, $r = \frac{kT}{\hbar}\exp(-E_{\mathrm{a}}/kT)$.
Then, the calculated rate at $T = 300 \, \mathrm{K}$ is $4.3 \times 10^{-3} \, \mathrm{s}^{-1}$,
which is much smaller than the calculated rate for thermal electron emission under thermal equilibrium.
Thus, the rate-determining step is suggested to be the migration of H$^{-}$ from Td site to BC site.
If the experimental activation energy, $E_{\mathrm{a}} = 0.84 \, \mathrm{eV}$, is used,
the calculated migration rate is $3.0 \times 10^{-1} \, \mathrm{s}^{-1}$, 
showing good agreement with the experimental rate, $2.8 \times 10^{-1} \, \mathrm{s}^{-1}$.
This fact strongly suggests that the activation energy observed in the DLTS experiment 
corresponds to the one for the migration process of H$^{-}$,
and the migration is the rate-limiting step for the thermal electron emission from H$^{-}$.
It should be noted here that the rate for the migration is estimated 
with treating hydrogen as a classical particle, thus neglecting nuclear quantum effect such as quantum tunneling.
Although the quantum tunneling can be dominant as migration mechanism at room temperature,
we expect that its effect is small because the distance between 
the Td site and the BC site is large as shown in Fig\@. \ref{fig:pes}.

\section{Conclusion}
Here, we study the thermal electron emission process of H$^{-}$ in silicon, 
$\mathrm{H}^{-} \to \mathrm{H}^{0} + e^{-}$, with density functional calculations.
Our calculation indicates that the electron emission process from H$^{-}$ in Si consists of two steps.
The first step is H$^{-}$ migration from its most stable Td site to a metastable BC site.
This migration is the rate-limiting step for the whole reaction with calculated activation energy of 0.96 eV.
The second step is the electron emission from H$^{-}$ in the BC site to a conduction band.
This step is much faster than the first step, having the calculated activation energy of 0.46 eV.
The activation energy observed in a DLTS experiment, 0.84 eV, is likely to correspond to the one in the first step.
Our study reveals that modern first-principles calculation can elucidate nonadiabatic process of hydrogen impurity, 
which is the most ubiquitous and mysterious impurity in semiconductors,
and thus shows the possibility to clarify the degradation mechanism of 
semiconductor devices by hydrogen impurity from first principles.

%\nocite{*}

\begin{acknowledgments}
The calculations were performed on the supercomputers at the Institute for Solid State Physics, the University of Tokyo. 
This research was supported by MEXT as “Priority Issue on Post-K computer” (Development of new fundamental technologies for
high-efficiency energy creation, conversion/storage and use) and 
by JSPS KAKENHI Grant-in-Aid for Scientific Research on Innovative Areas “Hydrogenomics”, No. JP18H05519.
We receive further support from a project commissioned by the New Energy and Industrial Technology Development Organization (NEDO).
\end{acknowledgments}

\bibliography{main}% Produces the bibliography via BibTeX.

\end{document}